\documentclass[11pt]{article}
\usepackage{moriond,epsfig,amsmath,amssymb,color}
\usepackage[bookmarks=false]{hyperref}
\pdfoutput=1
\allowdisplaybreaks

\def \refeq#1{(\ref{#1})}

\def \reffig#1{Figure~\ref{#1}}

\def \cLdB1{{{\cal L}_{\Delta B = 1}^{\rm EW}}} 

\def \Op{Q}

\def \One{\leavevmode\hbox{\small1\kern-3.6pt\normalsize1}} 

\def \TeV{{\rm \; TeV}}

\def \B0toK0mumu{{ \bar{B}^0 \to K^0 \bar{\mu} \mu}}

\def \beq{\begin{equation}}
\def \eeq{\end{equation}}
\def \bea{\begin{eqnarray}}
\def \eea{\end{eqnarray}}

\def\bm#1{\mbox{\boldmath$#1$\unboldmath}}

\begin{document}
\title{\begin{boldmath} (NO) NEW PHYSICS IN $B_{s}$ MIXING AND DECAY\end{boldmath}}

\author{Ulrich Haisch}

\address{Rudolf Peierls Centre for Theoretical Physics \\
    University of Oxford,  1 Keble Road,   OX1 3PN Oxford, United Kingdom}

\maketitle\abstracts{The status of possible new-physics signals  in $B_{s,d}$-meson mixing and decay is reviewed. In particular, it is emphasized that the recent LHCb results, that  find no evidence for a non-standard phase in  $B_s$--$\bar B_s$ 
mixing, make a consistent explanation of the D\O \ data on the like-sign dimuon charge asymmetry  notoriously difficult. In order to  clarify 
the inconclusive experimental situation, 
independent measurements of the semileptonic asymmetries are needed.
}

\section{Introduction}

The phenomenon of neutral $B_s$-meson mixing is encoded in the
off-diagonal elements $M_{12}^s$ and $\Gamma_{12}^s$ of the mass and
decay rate matrix. These two complex parameters can be fully
determined by measuring the mass difference $\Delta M_s = M_H^s -
M_L^s$, the CP-violating phase $\phi_{J/\psi \phi}^s = -2 \beta_s$,
the decay width difference $\Delta \Gamma_s = \Gamma_L^s -
\Gamma_H^s$, and the CP asymmetry $a_{fs}^s$ in flavor-specific
decays.

The combined Tevatron and LHC determination of the mass difference
reads\cite{Abulencia:2006ze,Aaij:2011qx} 
\beq \label{eq:DMsexp}
\Delta M_s = (17.73 \pm 0.05) \, {\rm ps}^{-1} \,,
\eeq
and agrees well with the corresponding standard model (SM) prediction\cite{Lenz:2012mb}
\beq \label{eq:DMsSM}
(\Delta M_s)_{\rm SM} =  (17.3 \pm 2.6) \, {\rm ps}^{-1} \,.
\eeq
Here the quoted errors correspond to 68\% confidence level (CL)
ranges.

The phase difference $\phi_{J/\psi\phi}^s$ between the $B_s$ mixing
and the $b \to s c \bar c$ decay amplitude and the width difference
$\Delta \Gamma_s$ can be simultaneously determined from an analysis of
the flavor-tagged time-dependent decay $B_s \to J/\psi \phi$.
Such measurements have been performed by CDF, D\O, and recently also by LHCb.
 Combining  $1\, {\rm fb}^{-1}$ of 
$B_s \to J/\psi \phi$ and $B_s \to J/\psi f_0$
 data the LHCb collaboration finds\cite{Clarke} 
\beq \label{eq:phiDGnew}
\phi_{J/\psi\phi}^s = \left (-0.11 \pm 5.0 \right)^\circ \,, \qquad \Delta
\Gamma_s = \left (0.116 \pm 0.019 \right ) {\rm ps}^{-1} \,.
\eeq
The corresponding numbers
in the SM are\cite{Lenz:2012mb}
\beq \label{eq:phiDGSM}
(\phi_{J/\psi\phi}^s)_{\rm SM} = \arg \left [ \frac{\left (V_{ts}^\ast
      V_{tb} \right )^2}{\left (V_{cs}^\ast V_{cb} \right )^2} \right
] = \left (-2.1 \pm 0.1 \right)^\circ \,, \qquad (\Delta
\Gamma_s)_{\rm SM} = \left (0.087 \pm 0.021 \right ) {\rm ps}^{-1} \,, 
\eeq
and again agree well with the observations with the results for $\Delta \Gamma_s$ differing by 
$1.0 \sigma$.

The CP asymmetry in flavor-specific decays $a_{fs}^s$ can be extracted
from a measurement of the like-sign dimuon charge asymmetry $A_{\rm
  SL}^b$, which involves a sample that is almost evenly composed of
$B_d$ and $B_s$ mesons. Employing $9.0 \, {\rm
  fb}^{-1}$ of data the D\O \ collaboration obtains\cite{Abazov:2011yk}
\beq \label{eq:ASLbnew}
A_{\rm SL}^b = (-7.87 \pm 1.96 )
\cdot 10 ^{-3}\,.
\eeq
Utilizing the SM predictions for the individual
flavor-specific CP asymmetries\cite{Lenz:2012mb}
\beq \label{eq:afsSM}
(a_{fs}^d)_{\rm SM} = -(4.1\pm 0.6) \cdot 10^{-4} \,, \qquad
(a_{fs}^s)_{\rm SM} = (1.9\pm 0.3) \cdot 10^{-5} \,,
\eeq
one obtains
\beq \label{eq:ASLbSMnew}
(A_{\rm SL}^b)_{\rm SM} = (0.595 \pm 0.022) \hspace{0.5mm}
(a_{fs}^d)_{\rm SM} + (0.406 \pm 0.022) \hspace{0.5mm} (a_{fs}^s)_{\rm
  SM} = (-2.4 \pm 0.4) \cdot 10^{-4} \,.
\eeq
Using the
measured value of the CP asymmetry in flavor-specific $B_d$ decays\cite{Asner:2010qj}
\beq \label{eq:afsd}
a_{fs}^d = \left (-4.7 \pm 4.6 \right ) \cdot 10^{-3} \,,
\eeq
and \refeq{eq:ASLbnew} in \refeq{eq:ASLbSMnew}, one can also directly
derive a value of $a_{fs}^s$. One arrives at
\beq \label{eq:afssnew}
a_{fs}^s = (-1.3 \pm 0.8) \cdot 10^{-2} \,,
\eeq
The results \refeq{eq:ASLbnew} and \refeq{eq:ASLbSMnew} correspond 
to  a tension with a statistical significance of $3.9 \sigma$, while 
\refeq{eq:afssnew} differs from  $(a_{fs}^s)_{\rm SM}$ as given in  
\refeq{eq:afsSM} by $1.5 \sigma$ only.

\section{Model-Independent Analysis}

In view of the observed departures from the SM predictions, it is
natural to ask what kind of new physics is able to simultaneously
explain the measured values of $\Delta M_s$, $\phi_{J/\psi \phi}^s$,
$\Delta \Gamma_s$, and $a_{fs}^s$.  This question can be addressed 
 in a model-independent way by parametrizing the off-diagonal 
elements of the mass and decay rate matrix as follows
\begin{gather} 
M_{12}^s = (M_{12}^s)_{\rm SM} + (M_{12}^s)_{\rm NP} = (M_{12}^s)_{\rm
SM} \, R_{M} \, e^{i \phi_{M}} \,, \nonumber \\[-1.5mm]
\label{eq:M12G12paraNP} \\[-1.5mm] \Gamma_{12}^s =
(\Gamma_{12}^s)_{\rm SM} + (\Gamma_{12}^s)_{\rm NP} =
(\Gamma_{12}^s)_{\rm SM} \, R_{\Gamma} \, e^{i \phi_{\Gamma}} \,.
\nonumber
\end{gather}
In he presence of generic new physics, the $B_s$-meson observables of
interest are then given to leading power in
$|\Gamma_{12}^s|/|M_{12}^s|$ by
\begin{gather} 
  \Delta M_s = (\Delta M_s)_{\rm SM} \hspace{0.5mm} R_{M} \,, \qquad
  \phi_{J/\psi \phi}^s = (\phi_{J/\psi \phi}^s)_{\rm SM} + \phi_{M}
  \,, \nonumber \\[1mm] \Delta \Gamma_s = (\Delta \Gamma_s)_{\rm SM}
  \hspace{0.25mm} R_{\Gamma} \, \frac{\cos \left (\phi_{\rm SM}^s +
      \phi_{M} - \phi_{\Gamma} \right )} {\cos \phi_{\rm SM}^s }
  \approx (\Delta \Gamma_s)_{\rm SM} \hspace{0.5mm} R_{\Gamma} \, \cos
  \left ( \phi_{M} - \phi_{\Gamma} \right )
  \,, \label{eq:observablesNP} \\[2mm] a_{fs}^s = (a_{fs}^s)_{\rm
    SM}\, \frac{R_{\Gamma}}{R_{M}} \; \frac{\sin \left (\phi_{\rm
        SM}^s + \phi_{M} - \phi_{\Gamma} \right )}{\sin \phi_{\rm
      SM}^s } \approx (a_{fs}^s)_{\rm SM}\, \frac{R_{\Gamma}}{R_{M}}
  \, \frac{\sin \left (\phi_{M} - \phi_{\Gamma} \right )}{ \phi_{\rm
      SM}^s} \,, \nonumber
\end{gather}
where the final results for  $\Delta \Gamma_s$ and $a_{fs}^s$ have been obtained by performing an expansion in $\phi_{\rm SM}^s = \arg
\left (-(M_{12}^s)_{\rm SM}/(\Gamma_{12}^s)_{\rm SM} \right )$. Since $\phi_{\rm
  SM}^s = (0.22 \pm 0.06)^\circ \;$\cite{Lenz:2012mb} this is 
an excellent approximation.

\begin{figure} 
\begin{center}
\includegraphics[width=6.75cm]{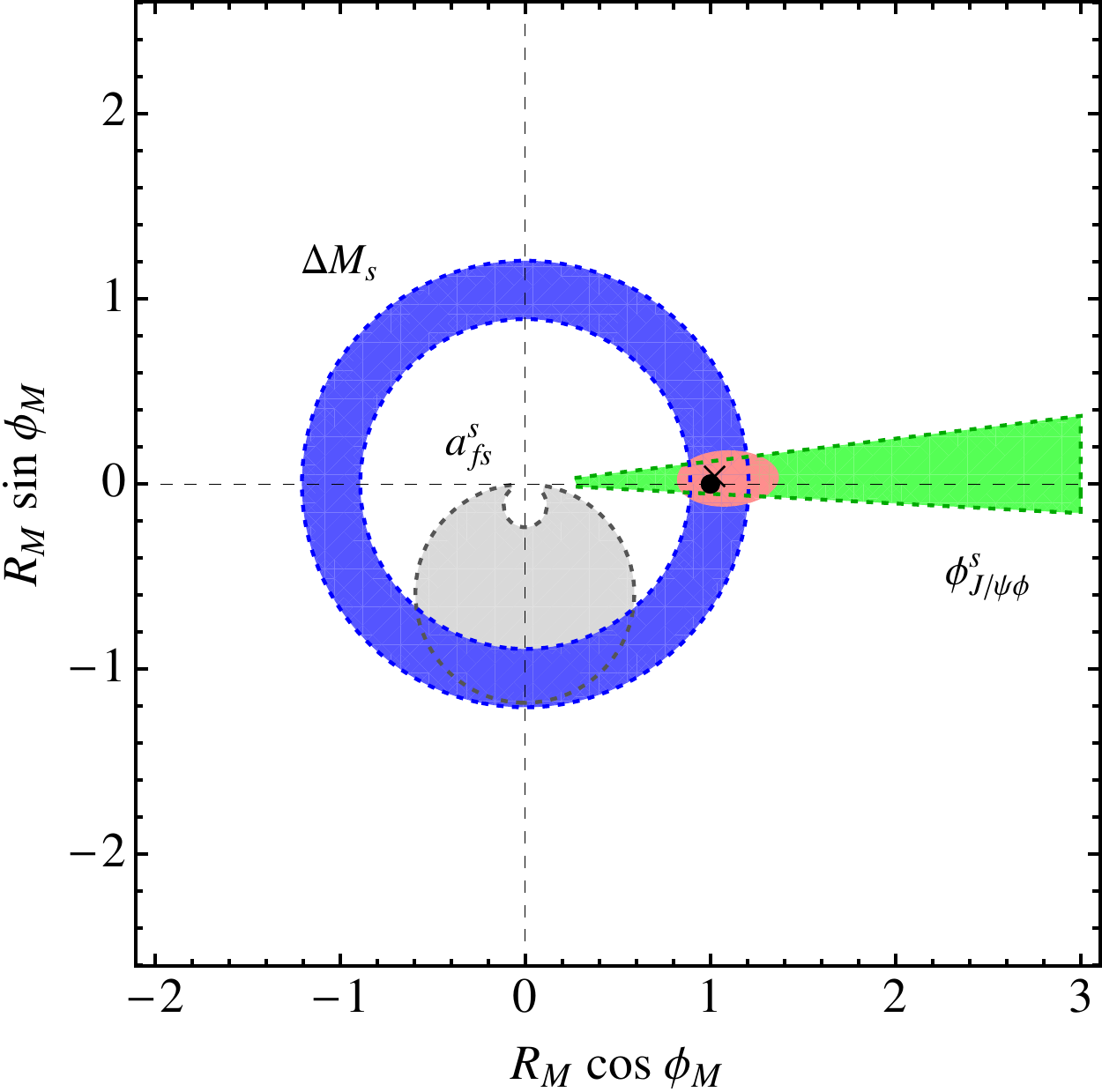} \qquad 
\includegraphics[width=6.75cm]{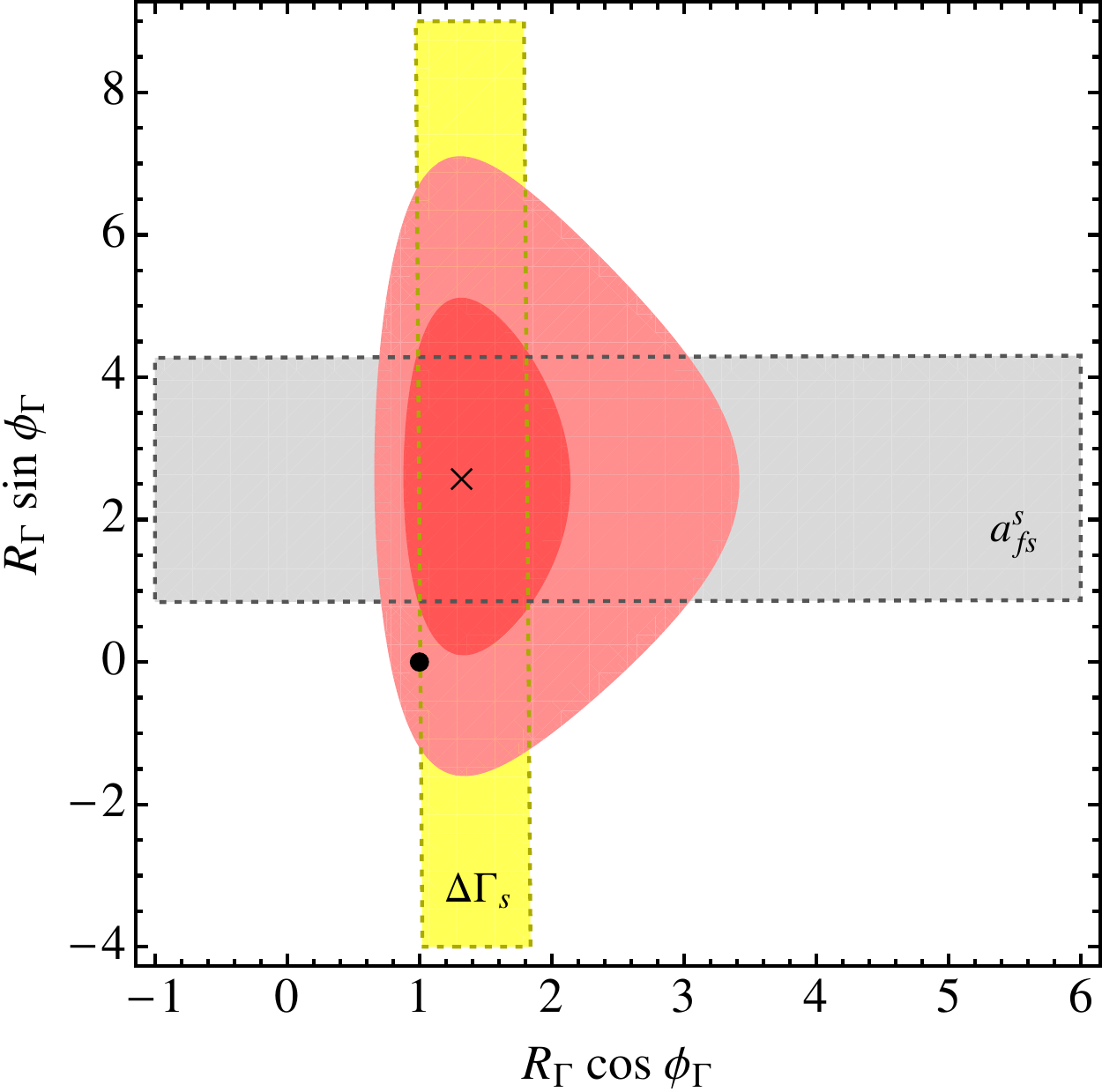}
\caption{\label{fig:fits} Left (Right):
      Constraints on  $R_M$ and $\phi_M$ ($R_\Gamma$ and
      $\phi_\Gamma$) in
      scenario S1 (S2). For the individual constraints the colored
      areas are  $68\%$ CL regions (${\rm dofs} = 1$), while for
      the combined fit  the red (light red)
      area is the $68\%$ ($95\%$) probability region (${\rm dofs} = 2$).
    The SM values (best-fit solutions) are marked
      by a dot (cross). \mbox{~~~}}
\end{center}
\end{figure}

The four real parameters $R_{M,\Gamma}$ and $\phi_{M,\Gamma}$ entering
\refeq{eq:M12G12paraNP} can be constrained by confronting the observed
values of $\Delta M_s$, $\phi_{J/\psi \phi}^s$, $\Delta \Gamma_s$, and
$a_{fs}^s$ with their SM predictions.   We begin our analysis by
asking how well the SM hypothesis describes the data. Performing a global fit, we obtain $\chi^2= 3.5$ corresponding to
$0.7\sigma$ ($1.4\sigma$) for 4 (2) degrees of freedom (dofs). For the best-fit solution, we find instead $(R_M, \phi_M, R_\Gamma, \phi_\Gamma) = (1.02, 
2^\circ, 2.9, 65^\circ)$. After marginalization the corresponding 
symmetrized 68\% CL parameter ranges are
 \beq \label{eq:D1CL68}
\begin{aligned}
  R_M & = 1.04 \pm 0.16 \,, & \qquad \phi_M & = (-0.4 \pm 5.0)^\circ \,, \\
  R_\Gamma & = 3.4 \pm 1.8 \,, & \qquad \phi_\Gamma & = (56 \pm
  22)^\circ \,.
\end{aligned}
\eeq
Focusing on $R_\Gamma$ and $\phi_G$, we see that a very good fit  requires excessive corrections to $\Gamma_{12}^s$. 

In order to further elucidate the latter point, we analyse two orthogonal
hypothesis of new physics in $B_s$--$\bar B_s$ oscillations, namely a
scenario with $(M_{12}^s)_{\rm NP} \neq 0$ and $(\Gamma_{12}^s)_{\rm
  NP} = 0$ and a scenario with $(M_{12}^s)_{\rm NP} = 0$ and
$(\Gamma_{12}^s)_{\rm NP} \neq 0$. The left (right) panel in
\reffig{fig:fits} shows the results of our global fit for
scenario S1 (S2) in the $R_M \hspace{0.25mm} \cos \phi_M$--$R_M
\hspace{0.25mm} \sin \phi_M$ ($R_\Gamma \hspace{0.25mm} \cos
\phi_\Gamma$--$R_\Gamma \hspace{0.25mm} \sin \phi_\Gamma$) plane.  From the
left panel, we glean that in the scenario S1 the regions of all
individual constraints apart from $a_{fs}^s$  overlap. Minimizing
the $\chi^2$ function gives $(R_M, \phi_M) = (1.03, 2.0^\circ)$ and
$\chi^2/{\rm dofs} = 3.4/2$ corresponding to $1.3 \sigma$, which
represents only a marginal improvement with respect to the SM
hypothesis.\cite{Lenz:2012mb,Bobeth:2011st} In consequence, the case for a non-zero non-standard
contribution to $M_{12}^s$ is rather weak.  By inspection of the right panel, we see that in contrast to S1, in the scenario S2 a description of the data with a
probability of better than 68\% is possible. The best-fit point is
located at $(R_\Gamma, \phi_\Gamma) = (2.9, 62^\circ)$. In fact, the
latter parameters lead to an almost perfect fit with $\chi^2/{\rm
  dofs} = 0.2/2$ corresponding to $0.1\sigma$. The data 
hence statistically favors the new-physics scenario S2 over the
hypothesis S1. 

\section{New Physics in $\bm{b \to s \tau^+ \tau^-}$}

The above  findings suggest that one hypothetical explanation of  the 
experimentally observed large negative values of $a_{fs}^s$ (or equivalent 
$A_{\rm SL}^b$) consists in postulating new physics in $\Gamma_{12}^s$ 
that changes the SM value by  a factor of  3 or more. In the following we will 
study whether or not and to which extend such an speculative 
option is viable. While in principle any composite operator $(\bar s b)
\hspace{0.25mm} f$, with $f$ leading to an arbitrary flavor neutral
final state of at least two fields and total mass below the
$B_s$-meson mass, can contribute to $\Gamma_{12}^s$, the field content
of $f$ is in practice very restricted, since $B_s \to f$ and $B_d \to
X_s\hspace{0.25mm} f$ decays to most final states involving light
particles are severely constrained. One notable exception is the
subclass of $B_s$- and $B_d$-meson decays to a pair of tau leptons, 
as has been first pointed out a few years ago.\cite{Dighe:2007gt}

The possibility of large  $b \to s \tau^+ \tau^-$ contributions to $\Gamma_{12}^s$, 
can be analyzed in a model-independent fashion\cite{Bobeth:2011st} by adding 
\beq \label{eq:Leffsbtautau}
{\cal L}_{\rm eff} =  \frac{4 G_F}{\sqrt 2} \, V_{ts}^\ast V_{tb} \, \sum_i C_i (\mu) \, \Op_i \,,
\eeq
to the effective $\Delta B = 1$ SM Lagrangian. 
Here  $\mu$ denotes the renormalization scale and  the Fermi constant  $G_F$ as well as the 
 leading Cabibbo-Kobayashi-Maskawa (CKM) factor $V_{ts}^\ast V_{tb}$ have been extracted as
 global prefactors. The index
$i$ runs over the complete set of dimension-six operators with flavor
content $(\bar s b)(\bar \tau \tau)$, namely ($A, B = L, R$)
\beq \label{eq:Qsbtautau}
\begin{split}
  \Op_{S, AB} & = \left (\bar s \, P_A \, b \right ) \left (\bar \tau
    \, P_B \, \tau \right ) \,, \\
  \Op_{V, AB} & = \left (\bar s \, \gamma^\mu P_A \, b \right )
  \left (\bar \tau\, \gamma_\mu P_B \, \tau \right ) \,, \\
  \Op_{T, A} & = \left (\bar s \, \sigma^{\mu\nu} P_A \, b \right )
  \left (\bar \tau \, \sigma_{\mu \nu} P_A \,\tau \right ) \,,
\end{split}
\eeq
where $P_{L,R} = (1 \mp
\gamma_5)/2$ project onto left- and right-handed chiral fields and
$\sigma^{\mu \nu} = i \left [\gamma^\mu, \gamma^\nu \right
]/2$. 

\begin{figure} 
\begin{center}
\includegraphics[width=10cm]{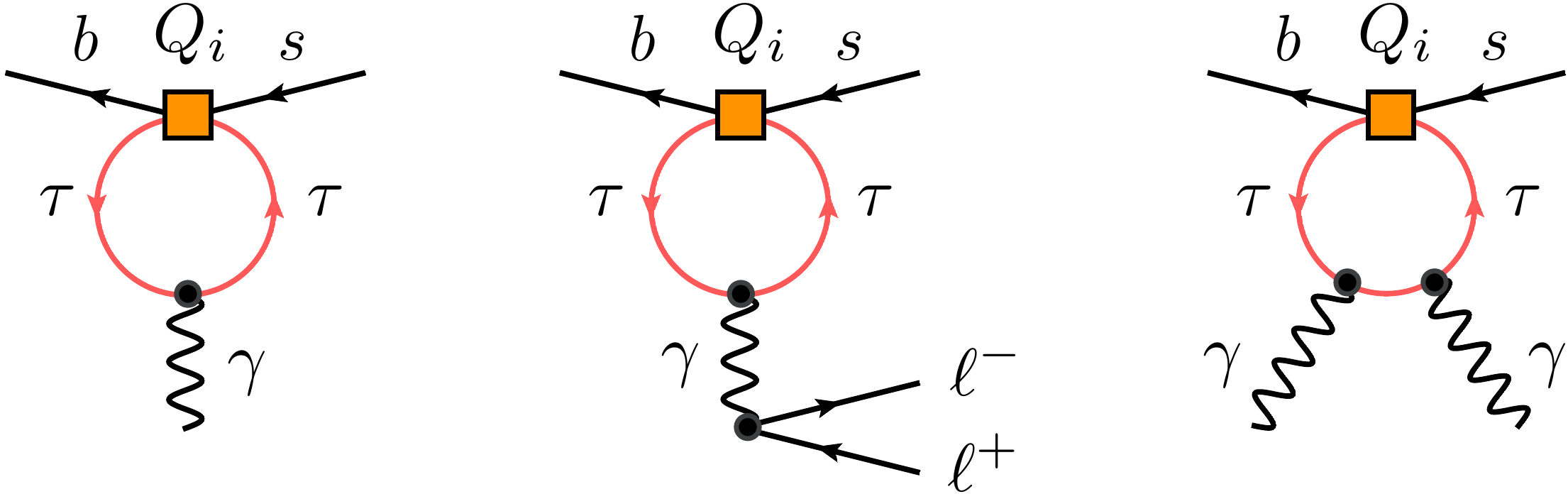} 
\caption{\label{fig:mixing} One-loop diagrams with a penguin
      insertion of a $(\bar s b) (\bar \tau \tau)$ operator (orange
      squares) that contribute to the processes $b\to s \gamma$ (left), $b \to s \ell^+ \ell^-$ (middle), and $b\to s \gamma\gamma$ (right). The tau loop in all graphs is closed.\mbox{~~~~}}
\end{center}
\end{figure}

The ten operators entering \refeq{eq:Qsbtautau} govern the purely
leptonic $B_s \to \tau^+ \tau^-$ decay, the inclusive semi-leptonic $B
\to X_s \tau^+\tau^-$ decay, and its exclusive counterpart $B^+ \to
K^+ \tau^+\tau^-$, making these channels potentially powerful
constraints. In practice, however, flavor-changing neutral current
$B_{s,d}$ decays into final states involving taus are experimentally
still largely unexplored territory, so that these direct constraints turn 
out to be not very strong. Explicitly, one obtains
\beq \label{eq:directBR}
\begin{split}
{\cal B}(B_s \to \tau^+\tau^-) \, & < \, 3\% \,, \\
{\cal B} (B \to X_s\tau^+\tau^-) \, & \lesssim \, 2.5 \% \,, \\
{\cal B} (B^+ \to K^+ \tau^+\tau^-) \, & < \, 3.3 \cdot 10^{-3} \,.
\end{split}
\eeq
Here the first limit derives\cite{Dighe:2010nj} from comparing the SM prediction $\tau_{B_s}/\tau_{B_d} - 1 \in 
[-0.4, 0.0] \%\;$\cite{Lenz:2012mb} with the corresponding experimental result
$\tau_{B_s}/\tau_{B_d} - 1 = (0.4 \pm 1.9) \%$, while the second (crude) bound follows from estimating\cite{Bobeth:2011st} the possible contamination of the exclusive and inclusive semileptonic decay samples by $B \to X_s \tau^+ \tau^-$ events. Limits on ${\cal B}(B_s \to \tau^+\tau^-)$ and ${\cal B} (B \to X_s\tau^+\tau^-)$ of strength similar to those given in \refeq{eq:directBR}
 also follow from charm counting\cite{Kagan:1997qna} and/or  LEP searches for $B$ decays with large missing energy.\cite{Grossman:1996qj} The final number corresponds to the 90\% CL upper limit on the branching ratio of $B^+ \to
K^+ \tau^+ \tau^-$ as measured by BaBar.\cite{Flood:2010zz}

Further constraints on the Wilson coefficients of the $(\bar s b)(\bar \tau \tau)$ operators 
arise indirectly from the experimentally available information on the
$b \to s \gamma$, $b \to s \ell^+ \ell^-$ ($\ell = e, \mu$), and $b \to s \gamma \gamma$.
transitions, because some of the effective operators introduced in
\refeq{eq:Qsbtautau} mix into the electromagnetic dipole operators $Q_{7,A}$ and the vector-like semileptonic operators $Q_{9,A}$. The relevant Feynman diagrams are shown in Figure~\ref{fig:mixing}. An explicit calculation\cite{Bobeth:2011st} shows that the operators
$\Op_{S,AB}$ mix neither into $\Op_{7,A}$ nor
$\Op_{9,A}$, while $\Op_{V,AB}$ ($\Op_{T,A}$)
mixes only into $Q_{9,A}$ ($\Op_{7,A}$). As a result of the particular mixing pattern, the
stringent constraints from the radiative decay $B \to X_s \gamma$ rule
out large contributions to $\Gamma_{12}^s$ only if they arise from the
tensor operators $\Op_{T,A}$.  Similarly, the rare decays $B \to X_s
\ell^+ \ell^-$ and $B \to K^{(\ast)} \ell^+ \ell^-$ primarily limit
contributions stemming from the vector operators $\Op_{V,AB}$. In contrast to $B \to X_s \gamma$, all $(\bar s b) (\bar \tau \tau)$
operators contribute to the double-radiative $B_s \to \gamma \gamma$ decay at
the one-loop level. A detailed study\cite{Bobeth:2011st} shows however  that the limits following from $b \to s \gamma \gamma$ are in practice not competitive with the bounds obtained  from the other tree- and loop-level mediated $B_{s,d}$-meson decays. 

\begin{table}
\begin{center}
\begin{tabular}{|c|c|c|c|}
  \hline
  Operator &   Bound on $C_i (m_b)$  & Bound on $\Lambda$ & Observable \\
  \hline
  $(\bar s \, P_A \, b) (\bar \tau \, P_B \, \tau)$ & 0.5 & $2.0 \TeV$  & $B_s \to  \tau^+ \tau^-$ \\
  $(\bar s \, \gamma^\mu P_A \, b) (\bar \tau \, \gamma_\mu P_B \, 
  \tau)$ &  0.8 & $1.0 \TeV$  & $B^+ \to K^+ \tau^+ \tau^-$ \\
  $(\bar s \, \sigma^{\mu \nu} P_L \, b) (\bar \tau \, \sigma_{\mu \nu} 
  P_L \, \tau)$ & 0.06 & 
  $3.2 \TeV$ & $B \to X_s \gamma$ \\
  $(\bar s \, \sigma^{\mu \nu} P_R \, b) (\bar \tau \, \sigma_{\mu \nu} 
  P_R \, \tau)$ & 0.09 & $2.8 \TeV$ & $B \to X_s \gamma$ \\
  \hline
\end{tabular}

\vspace{4mm}

  \parbox{15.5cm}{
    \caption{\label{tab:bounds} Model-independent limits on the
      Wilson coefficients of $(\bar s b) (\bar \tau \tau)$ operators. The second (third) column shows the limit on $C_i(m_b)$
 (the bound on the suppression scale $\Lambda$
      assuming a coupling strength of 1).}}
  \end{center}
\end{table}

\begin{figure} 
\begin{center}
\includegraphics[width=6.75cm]{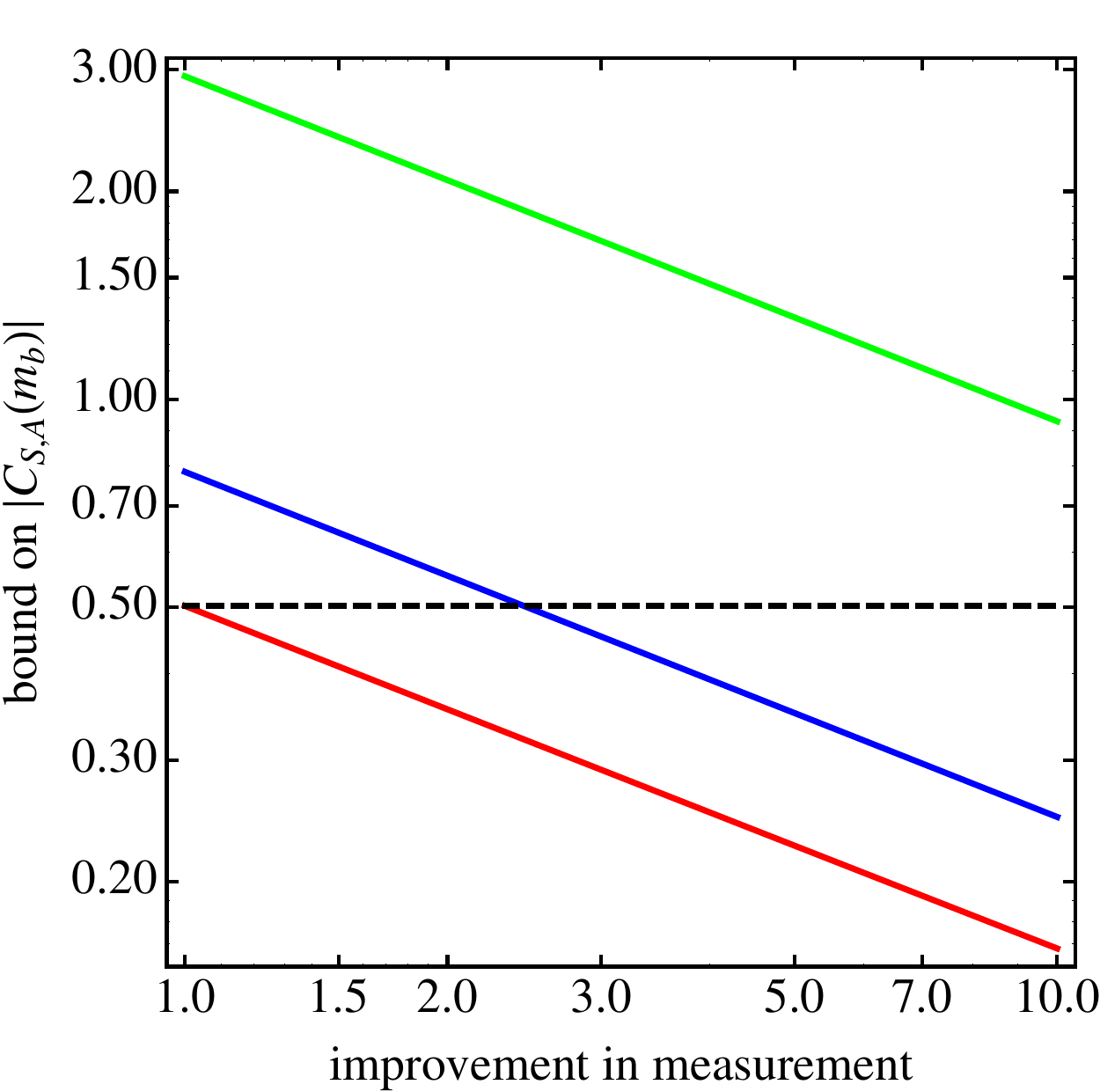} \qquad 
\includegraphics[width=6.75cm]{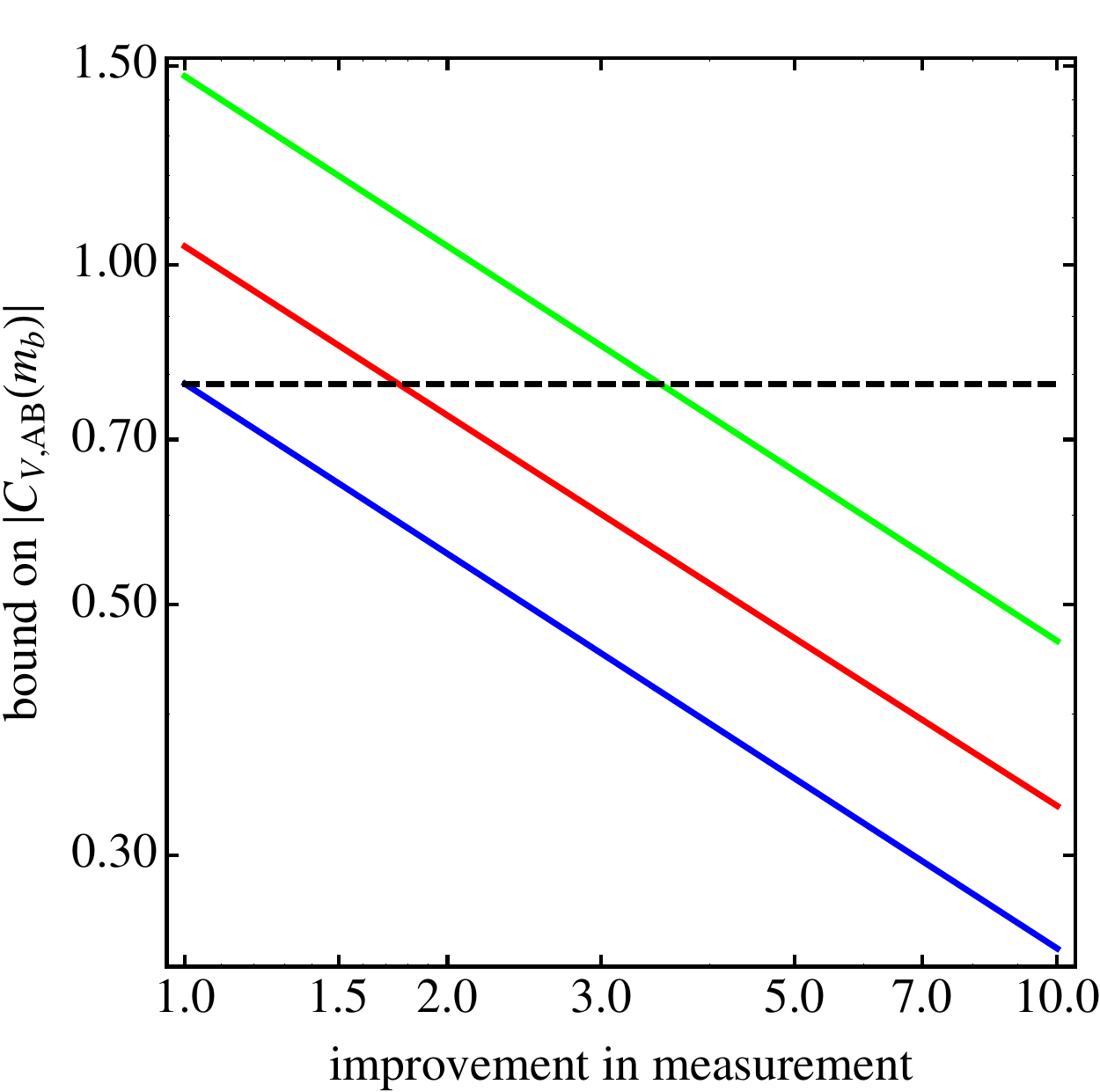}
\caption{\label{fig:comparison} Left (Right): Bound on $C_{S,A} (m_b)$ ($C_{V,AB} (m_b)$) from improved determinations of  ${\cal B} (B_s \to \tau^+ \tau^-)$ (red), ${\cal B} (B \to X_s \tau^+ \tau^-)$ (green), and ${\cal B} (B^+ \to K^+ \tau^+ \tau^-)$ (blue). The dashed lines indicate the presently best bound. \mbox{~}}
\end{center}
\end{figure}

The model-independent 90\% CL limits on the magnitudes of the Wilson coefficients are summarized in Table~\ref{tab:bounds}.
We see that in the case of the $(\bar s b)(\bar \tau \tau)$  scalar and vector  operators the allowed effects can reach almost ${\cal O} (1)$. These Wilson coefficients can hence be similar in size   to that of  the color-singlet current-current operator $Q_2$, which provides the dominant contribution to $\Gamma_{12}^s$ in the SM. In consequence, the corresponding suppression scale $\Lambda$ is quite low, around $(1-2) \, {\rm TeV}$. Possible new-physics contributions to the $(\bar s b)(\bar \tau \tau)$ tensor operators are more severely 
constrained, because they lead to a contamination of $B \to X_s \gamma$. It is also interesting to ask which impact future improved extractions of ${\cal B} (B_s \to \tau^+ \tau^-)$, ${\cal B} (B \to X_s \tau^+ \tau^-)$, and ${\cal B} (B^+ \to K^+ \tau^+ \tau^-)$ will have on the limits on the Wilson coefficients $C_i(m_b)$. Such a comparison is provided in Figure~\ref{fig:comparison}. From the left panel one concludes that  ${\cal B} (B^+ \to K^+ \tau^+ \tau^-) < 1.3 \cdot 10^{-3}$, corresponding to an improvement of the present upper limit by a factor of 2.5, would allow to set a bound on $C_{S,A} (m_b)$ that is as good as the one that follows  at present already from $B_s \to \tau^+ \tau^-$. As can seen from the right panel, in the case of $C_{V,AB} (m_b)$,  limits of  ${\cal B} (B_s \to \tau^+ \tau^-) < 1.7\%$ and  ${\cal B} (B \to X_s \tau^+ \tau^-) < 0.7\%$ are needed to be competitive with $B^+ \to K^+ \tau^+ \tau^-$. The quoted limits correspond to  improvements of our current knowledge of the relevant $B$-meson branching ratios by a factor of  1.8 and 3.6, respectively.  Finally, for what concerns  $C_{T,A} (m_b)$, even improvements of the direct constraints by a factor of more than 10 are not sufficient to beat the indirect constraint arising from $B \to X_s \gamma$. 

\section{Effects of $\bm{(\bar s b)(\bar \tau \tau)}$ Operators in $\bm{\Gamma_{12}^s}$}

The off-diagonal element of the decay-width matrix is related via the
optical theorem to the absorptive part of the forward-scattering
amplitude which converts a $\bar B_s$ into a $B_s$ meson. Working to
leading order in the strong coupling constant and $\Lambda_{\rm QCD}/m_b$, the
contributions from the complete set of operators \refeq{eq:Qsbtautau}
to $\Gamma_{12}^s$ is found by computing the matrix elements of the $(Q_i, Q_j)$ double insertions 
between quark states. Such a calculation\cite{Bobeth:2011st} leads to the results 
\beq \label{eq:RGammaSVT}
\begin{split} 
  (R_\Gamma)_{S,AB} & < 1 + \left ( 0.4 \pm 0.1 \right )
  |C_{S,AB} (m_b)|^2 \,, \\[2mm]
  (R_\Gamma)_{V,AB} & < 1+ \left ( 0.4 \pm 0.1 \right )
  |C_{V,AB} (m_b)|^2 \,,\\[2mm]
  (R_\Gamma)_{T,A} & < 1+ \left ( 3.6 \pm 0.9 \right ) |C_{T,A}
  (m_b)|^2 \,,
\end{split}
\eeq
where the quoted uncertainties are due to the error on $(\Delta
\Gamma_s)_{\rm SM}$ as given in \refeq{eq:phiDGSM}. Employing now the
90\% CL bounds on the low-energy Wilson coefficients given in Table~\ref{tab:bounds},
it follows that
\beq \label{eq:RGammaSVTbounds} 
(R_\Gamma)_{S,AB} < 1.15 \,,  \qquad (R_\Gamma)_{V,AB} < 1.35 \,, \qquad
(R_\Gamma)_{T,L} < 1.02 \,, \qquad (R_\Gamma)_{T,R} < 1.04 \,.
\eeq
These numbers imply that $(\bar s b) (\bar \tau \tau)$ operators of
scalar (vector) type can lead to enhancements of $|\Gamma_{12}^s|$
over its SM value by 15\% (35\%) without violating any existing
constraint. In contrast, contributions from tensor operators can alter
$|\Gamma_{12}^s|$ by at most $4\%$.  These numbers should be compared to the best-fit solution for $R_\Gamma$ as given in \refeq{eq:D1CL68}. From the comparison it immediately becomes apparent  that absorptive new physics in 
$\Gamma_{12}^s$ in form of $(\bar s b)(\bar \tau \tau)$ operators cannot provide 
a satisfactory explanation of the anomaly in the dimuon charge asymmetry data 
observed by the D\O \ collaboration. This is a model-independent conclusion that can be shown to hold in explicit models 
of new physics with modification of the $b \to s \tau^+ \tau^-$ channel such as leptoquark scenarios or $Z^\prime$ models.\cite{Bobeth:2011st}

\section{Conclusions and Outlook}

Motivated by the observation that the anomalously large dimuon charge
asymmetry measured by the D\O \ collaboration, can be fully explained
only if new physics contributes to the absorptive part of the
$B_{s,d}$--$\bar B_{s,d}$ mixing amplitudes, we have presented a
model-independent study of the contributions to $\Gamma_{12}^s$
arising from the complete set of dimension-six $(\bar s b) (\bar \tau \tau)$ operators.  Taking into account the direct
bounds from $B_s \to \tau^+ \tau^-$, $B \to X_s \tau^+ \tau^-$, and
$B^+ \to K^+ \tau^+ \tau^-$ as well as the indirect constraints from
$b \to s \gamma$, $b \to s \ell^+ \ell^-$, and $b
\to s \gamma \gamma$, we have demonstrated that only the Wilson
coefficients of the tensor operators are severely constrained by data,
while those of the scalar and vector operators can be sizable and
almost reach the size of the Wilson coefficient of the leading
current-current SM operator. It
follows that the presence of a single $(\bar s b) (\bar \tau \tau)$
operator can lead to an enhancement of $\Gamma_{12}^s$ of at most 35\%
compared to its SM value. Since a resolution of 
the tension in the $B_{s}$-meson sector would require the effects to be 
of the order of 300\% (or larger), the allowed shifts are by far too small to 
provide an satisfactory explanation of the issue. We emphasize that after 
minor modifications, our general results can be applied to other 
dimension-six operators involving quarks and leptons. For example, 
as a result of the 90\%~CL limit ${\cal B} (B^+ \to K^+ \tau^\pm \mu^\mp) 
< 7.7 \cdot 10^{-5}$,\cite{Aubert:2007rn} the direct bounds on the Wilson
coefficients of the set of $(\bar s b) (\bar \tau \mu)$ operators turn
out to be roughly a factor of $7$ stronger than those in the $(\bar s
b) (\bar \tau \tau)$ case.  Possible effects of $(\bar s b) (\bar \tau
\mu)$ operators are therefore generically too small to lead to a
notable improvement of the tension present in the current $B_{s,d}$-meson data. 
 Similarly, a contribution from $(\bar d b) (\bar \tau \tau)$ operators to 
$\Gamma_{12}^d$ large enough to explain the $A_{\rm SL}^b$ data is 
excluded by  the 90\% CL bound ${\cal B} (B \to \tau^+ \tau^-) < 4.1 \cdot 
10^{-3}$.\cite{Aubert:2005qw} Naively, also $(\bar b d) (\bar c c)$ operators are 
heavily constrained (meaning that their Wilson coefficients should be 
smaller than those of the QCD/electroweak penguins in the  SM) by 
the plethora of exclusive $B$ decays. A dedicated analysis of the latter class of 
contributions is however not available in the literature.

Our model-independent study of non-standard effects in $\Gamma_{12}^s$
can readily be applied to explicit SM extensions involving leptoquarks or $Z^\prime$ bosons.  In fact, the pattern of
deviations found in these scenarios resembles the one 
of all new-physics model with real $r_{\rm NP} =  (M_{12}^s)_{\rm
  NP}/(\Gamma_{12}^s)_{\rm NP}$, for which it can be shown that the
measurement of $\Delta M_s$ generically puts stringent constraints on
both $\Delta \Gamma_s$ and $a_{fs}^s$. These bounds turn out to be 
weakest if the ratio $r_{\rm NP}$ is positive and as small as possible. Since
on dimensional grounds $r_{\rm NP}$ scales as the square of the
new-physics scale, this general observation implies that SM  extensions that aim at a good description of the Tevatron
data should have new degrees of freedom below the electroweak scale
and/or be equipped with a mechanism
that renders the contribution to $M_{12}^s$ small. Furthermore, models
in which $(M_{12}^s)_{\rm NP}$ is generated beyond Born level seem
more promising, since in such a case $r_{\rm NP}$ is suppressed by a
loop factor with respect to the case where $(M_{12}^s)_{\rm NP}$
arises already at tree level.

The above discussion implies that  a full explanation of the observed discrepancies is not even
possible for the most general case $(M_{12}^s)_{\rm NP} \neq 0$ and
$(\Gamma_{12}^s)_{\rm NP} \neq 0$. Numerically, one finds
that the addition of a single $(\bar s b) (\bar \tau \tau)$ vector
operator giving $(R_\Gamma)_{V,AB} = 1.35$ on top of dispersive new physics with $(M_{12}^s)_{\rm NP} \neq
0$, can only improve the quality of the fit to the latest set of
measurements to $\chi^2 = 1.4$ compared to $\chi^2 = 3.5$
within the SM.  This might indicate that the high central value of
$A_{\rm SL}^b$ observed by the D\O \ collaboration is (partly) due to
a statistical fluctuation.
Future improvements in the measurement of
the CP phase $\phi_{J/\psi \phi}^s$ and, in particular, a first
determination of the difference $a_{fs}^s-a_{fs}^d$ between the $B_s$
and $B_d$ semileptonic asymmetries by LHCb, are soon expected to shed
light on this issue, and are of utmost importance in order to answer whether or not there is  new physics hiding in the $B_{s,d}$-meson sector.

\section*{Acknowledgments}

A big ``thank you'' to Christoph~Bobeth  for a fruitful collaboration that forms the basis of this proceeding. 
I am also grateful to the organizers of Moriond Electroweak 2012 for the invitation to a great conference, and to Martin~Bauer,  Henning~Flaecher, Sabine~Kraml, Alex~Lenz, Nazila~Mahmoudi, Matthias~Neubert,  Jonas~Rademaker, Lisa~Randall, David~Straub, and Jure~Zupan (as well as those participants that I talked to, but that have escaped my mind) for numerous discussions on and off the slopes. Travel support from the UNILHC network (PITN-GA-2009-237920) is acknowledged.

\end{document}